\documentclass[sigconf]{acmart}

\AtBeginDocument{%
  \providecommand\BibTeX{{%
    \normalfont B\kern-0.5em{\scshape i\kern-0.25em b}\kern-0.8em\TeX}}}

\setcopyright{acmcopyright}
\copyrightyear{2018}
\acmYear{2018}
\acmDOI{10.1145/1122445.1122456}

\setcopyright{none}
\renewcommand\footnotetextcopyrightpermission[1]{}

\usepackage{algorithm}
\usepackage{algcompatible}
\usepackage{algpseudocode}
\usepackage{multirow} 
\usepackage{amsmath} 
\usepackage{xcolor}
\usepackage{color}
\usepackage{threeparttable}
\usepackage{amsfonts}
\usepackage{diagbox}
\usepackage{graphics}
\usepackage{epsfig}
\usepackage{xspace}
\usepackage{bm}
\usepackage{listings}
\usepackage{eucal}
\usepackage{url} 
\usepackage{adjustbox}
\usepackage{subfigure}
\usepackage{graphicx}
\usepackage[many]{tcolorbox}	
\usepackage{enumitem,kantlipsum}

\newcommand{\toolname}{\textit{InPaFer}\xspace}
\newcommand{\manualFix}{ManuallyFix\xspace}
\newcommand{\fixwithPathes}{FixWithPatches\xspace}
\newcommand{\fixwithtool}{FixWith\toolname}

\newtcolorbox{shadedbox}{
	drop shadow southeast,
	breakable,
	enhanced jigsaw,
	colback=white,
}

\newcounter{finding}
\newcommand{\finding}[1]{\refstepcounter{finding}
	\begin{shadedbox}
		\textbf{Finding \arabic{finding}. } \emph{#1}
	\end{shadedbox}
}
\newcommand{\method}{{\texttt{Modified Method}}\xspace}
\newcommand{\trace}{{\texttt{Execution Trace}}\xspace}
\newcommand{\variable}{{\texttt{Variable Value}}\xspace}

\begin{document}

\lstdefinestyle{java}{ 
	language=java,
	basicstyle=\footnotesize\ttfamily, 
	breakatwhitespace=false, 
	breaklines=true, 
	captionpos=b, 
	commentstyle=\color[rgb]{0,0.6,0}, 
	deletekeywords={}, 
	firstnumber=1, 
	frame=shadowbox, 
	frameround=tttt, 
	keywordstyle={[1]\color{blue!90!black}},
	keywordstyle={[3]\color{red!80!orange}},
	escapeinside={<@}{@>},
	morekeywords={}, 
	numbers=left, 
	numbersep=-5pt, 
	numberstyle=\tiny\color[rgb]{0.1,0.1,0.1}, 
	rulecolor=\color{black}, 
	showstringspaces=false, 
	showtabs=false, 
	stepnumber=1, 
	tabsize=2, 
	backgroundcolor=\color{white},
	linewidth=0.98\columnwidth
}

\title{Interactive Patch Filtering as Debugging Aid}

\author{Jingjing Liang}
\affiliation{
  \institution{Key Lab of High Confidence Software Technologies, Ministry of Education \\
  Department of Computer Science and Technology, EECS, Peking University}
  \city{Beijing}
  \country{China}            
}
\email{jingjingliang@pku.edu.cn}          

\author{Ruyi Ji}
\affiliation{
  \institution{Key Lab of High Confidence Software Technologies, Ministry of Education \\
  Department of Computer Science and Technology, EECS, Peking University}
  \city{Beijing}
  \country{China}            
}
\email{jiruyi910387714@pku.edu.cn}          

\author{Jiajun Jiang}
\affiliation{
  \institution{Key Lab of High Confidence Software Technologies, Ministry of Education \\
  Department of Computer Science and Technology, EECS, Peking University}
  \city{Beijing}
  \country{China}            
}
\email{jiajun.jiang@pku.edu.cn}          

\author{Yiling Lou}
\affiliation{
  \institution{Key Lab of High Confidence Software Technologies, Ministry of Education \\
  Department of Computer Science and Technology, EECS, Peking University}
  \city{Beijing}
  \country{China}            
}
\email{louyiling@pku.edu.cn}          

\author{Yingfei Xiong}
\affiliation{
  \institution{Key Lab of High Confidence Software Technologies, Ministry of Education \\
  Department of Computer Science and Technology, EECS, Peking University}
  \city{Beijing}
  \country{China}            
}
\email{xiongyf@pku.edu.cn}          

\author{Gang Huang}
\affiliation{
  \institution{Key Lab of High Confidence Software Technologies, Ministry of Education \\
  Department of Computer Science and Technology, EECS, Peking University}
  \city{Beijing}
  \country{China}            
}
\email{hg@pku.edu.cn}          

\begin{abstract}
	It is widely recognized that program repair tools need to have a high precision to be useful, i.e., the generated patches need to have a high probability to be correct. However, it is fundamentally difficult to ensure the correctness of the patches, and many tools compromise other aspects of repair performance such as recall for an acceptable precision. 

	In this paper we ask a question: can a repair tool with a low precision be still useful? To explore this question, we propose an interactive filtering approach to patch review, which filters out incorrect patches by asking questions to the developers. Our intuition is that incorrect patches can still help understand the bug. With proper tool support, the benefit outweighs the cost even if there are many incorrect patches. 

	We implemented the approach as an Eclipse
	plugin tool, \toolname, and evaluated it with 
	a simulated experiment and a user study with 30 developers. The results show that our approach improve the repair performance of developers, with 62.5\% more successfully repaired bugs and 25.3\% less debugging time in average. In particular, even if the generated patches are all incorrect, the performance of the developers would not be significantly reduced, and could be improved when some patches provide useful information for repairing, such as the faulty location and a partial fix.	
\end{abstract}



\maketitle

\renewcommand{\algorithmicrequire}{\textbf{Input:}} 
\renewcommand{\algorithmicensure}{\textbf{Output:}} 

\newcommand{\code}[1]{\texttt{#1}}
\newcommand{\todo}[1]{{\color{red} \bf \{TODO: {#1}\}}}

\def\modify#1#2#3{{\small\underline{\sf{#1}}:} {\color{red}{\small #2}}
{{\color{red}\mbox{$\Rightarrow$}}} {\color{blue}{#3}}}

\newcommand{\yxmodify}[2]{\modify{Yingfei}{#1}{#2}}
\newcommand\mymargin[1]{\marginpar{{\flushleft\textsc\footnotesize {#1}}}}
\newcommand\yxmargin[1]{\mymargin{YX:\;#1}}

\newcommand{\jiajun}[1]{{\color{orange}{[Jiajun: #1]}}}

\newcommand{\figref}[1]{Figure~\ref{#1}}
\newcommand{\tabref}[1]{Table~\ref{#1}}
\newcommand{\eqnref}[1]{violation~\eqref{#1}}
\newcommand{\secref}[1]{Section~\ref{#1}}
\newcommand{\tblref}[1]{Table~\ref{#1}}
\newcommand{\smalltitle}[1]{{\smallskip \noindent \bf  {#1}.\ }}
\newcommand{\smalltitlecolon}[1]{{\smallskip \noindent \bf  {#1}:\ }}

\newcommand{\yxmodifyok}[2]{#2}

\newcommand{\mycomment}[2]{{\small\color{magenta}\underline{\sf{#1}}:} {\color{magenta}{\small #2}}}
\newcommand{\yxcomment}[1]{\mycomment{Yingfei}{#1}}

\newcommand{\mycommentblue}[2]{{\small\color{red}\underline{\sf{#1}}:} {\color{red}{\small #2}}}
\newcommand{\ljj}[1]{\mycommentblue{ljj}{#1}}


\renewcommand{\COMMENT}[2][.5\linewidth]{%
	\leavevmode\hfill\makebox[#1][l]{{\footnotesize\ttfamily\textcolor{blue}{//~#2}}}}

\section{Introduction} \label{section:introduction_v1}

In the past decades, automatic program repair (APR) attracted a lot of research efforts and significant progress has been made~\cite{repair-survey2017,survey,kPAR-url,liu2019avatar,Wen2018ContextAwarePG,ISSTA18-SimFix,Xuan2016History,xiong-icse17,xuan2016nopol}. Many of the proposed APR approaches are test-based program repair approaches, which take as input a buggy program and a test suite with at least a failed test, and automatically generate a set of patches that make all tests pass. 

It is commonly recognized that an APR tool needs to have a high precision to be useful, i.e., the generated patches should have a high probability to be correct. For example, Tao et al.~\cite{Tao:2014:AGP:2635868.2635873} have shown that the repair performance of developers significantly improves when they are given a high-quality patch, and becomes worse when they are given a low-quality patch. If a repair approach produces more incorrect patches than correct patches, the overall repair performance would become even lower.

However, it is fundamentally difficult to achieve a high precision. Most programs are not developed with a formal and complete specification, and we usually only have a test suite as a specification. Since tests cannot guarantee the correctness of the program, it is therefore fundamentally impossible to guarantee the correctness of patches. Furthermore, in practice the test suites are usually weak, such that many more incorrect patches can pass the tests than correct patches~\cite{long2016analysis}, making achieving a high precision very difficult. This problem is often known as ``weak test suite''~\cite{PatchPlausibility} or ``overfitting''~\cite{smith2015cure}. 

As a result, many approaches take conservative strategies~\cite{DBLP:journals/chinaf/RoychoudhuryX19}, compromising performance in other aspects to reach an acceptable precision. For example, many APR tools return only one most probable patch for each bug~\cite{xiong-icse17,Wen2018ContextAwarePG,ISSTA18-SimFix,ssFix,ELIXIR}.
 Some approaches repair only bugs satisfying strict conditions, e.g., frequently recurring bugs~\cite{genesis,bader2019getafix} or bugs with a reference implementation~\cite{DBLP:conf/icse/MechtaevNNGR18,SearchRepair}.
 These conservative strategies inevitably compromise the performance in other aspects. 
 In the above examples, returning more patches or targeting more bugs could potentially help generate correct patches for more bugs, i.e., increasing recall, but the conservative strategies disallow this possibility.
 In fact, all state-of-the-art repair approaches have a low recall, i.e., only a small portion of bugs can be correctly fixed. For example, Hercules~\cite{DBLP:conf/icse/SahaSP19}, one of the newest repair approach proposed in 2019, repairs only 13\% bugs on the Defects4J benchmark. 

Given the fundamental difficulty in achieving a high precision, in this paper we explore from a different perspective: can a repair tool with a low precision still be useful? Our exploration is based on the following observation: The practical use of ARP tools consists of two steps~\cite{DBLP:conf/issta/FryLW12}, patch generation, where the ARP tool proposes candidate patches, and patch review, where the developer examines the patches to ensure their correctness and other quality attributes. While the patch generation has been extensively studied, we still lack understanding and tool support for patch review. We conjecture that the review of incorrect patches also helps understand the bug, and with proper tool support, the review of incorrect patches would at least not reduce the repair performance of the developers. If this conjecture holds, we have the liberty to build ARP tools with lower precisions, e.g., by generating more patches per bugs, and potentially increase other aspects of tool performance, e.g., recall.

We propose an interactive patch filtering approach to provide tool support for patch review, verifying the conjecture constructively.
Given multiple patches for a bug, where most of the patches are expected to be incorrect, an interactive patch filtering tool asks the developer questions about attributes of the system that could distinguish between different patches. 
The attributes could be about the behavior of a test, e.g., whether a statement should be executed during a test or not, or could be the confidence of the developer on the system, e.g., whether a method should be patched or not. 
The developer picks a question and provides an answer, and the tool filters the patches based on the answer. 
The process continues until the developer figures out a correct patch (by picking an automatically generated one or by contriving a new one) or no more patches can be filtered out. 
We design a two-stage algorithm to implement this interactive system, introducing an offline preparing stage to optimize the response time of the online interactive stage. 

We have implemented our approach as an Eclipse plugin called \toolname, which stands for an \textbf{In}teractive \textbf{Pa}tch \textbf{F}ilt\textbf{er}. 
The plugin includes a user interface to allow the developer to easily browse the patches and the questions, as well as a diff view to visualize the effect of a patch on a test execution. 
The plugin implements three types of questions: (1) whether a statement should be executed in a test execution, (2) whether the assignment to a variable is correct in a test execution, (3) whether a method should be patched or not.

Based on the plugin, we conducted two experiments to verify the conjecture. 
We collected the patches generated by 13 different repair tools on the Defects4J benchmark~\cite{just2014defects4j} and assume that the patches are generated by one tool. 
In this way, we get a combined tool whose precision is lower than every single tool but the recall is higher than every single tool. 
Based on this tool, the first experiment is a simulated experiment that measures the average number of questions to distinguish the patches, where a computer-simulated user randomly chooses questions and provides answers. The result suggests that a relatively small number of questions are needed to finish the filtering process, 3.1 per bug on average, and the number of needed questions is not related to the number of patches. 

The second experiment is a user study involving 30 participants divided into three groups. 
The three groups repair bugs without any patch, with the generated patches, and with \toolname, respectively. 
The results show that, compared with the group with no patch, the group with \toolname correctly repair 63\% more bugs and uses 25\% less time on average; compared with the group with the generated patches, the group with \toolname correctly repair 39\% more bugs and uses 28\% less time on average. 
Furthermore, even if all generated patches are incorrect, the group with \toolname still performs slightly better than the group with no patches, repairing 27\% more bugs and uses 13\% less time. This confirms our conjecture: with proper tool support, the patch review process helps understand the bug, and eventually contribute to debugging.  

In summary, this paper makes the following contributions:
\begin{enumerate}
	\item An interactive patch filtering approach to supporting the patch review step, and a two-stage algorithm to implement the approach. 
	\item An Eclipse plugin with a carefully designed interface for the user to easily browse the questions and the patches. 
	\item Two experiments, including a user study, to investigate the usefulness of \toolname in aiding developers in debugging.
\end{enumerate}

The remainder of the paper is organized as follows. 
Section~\ref{section:approach} introduces the framework of our approach. 
Section~\ref{section:implementation} illustrates our implementation in detail. 
Section~\ref{section:evaluation} evaluates the effectiveness of our approach on Defects4J, while Section~\ref{section:threats} and ~\ref{section:relatedwork} validates the threats and related work, respectively. 
Finally, Section~\ref{section:conclusion} concludes the paper.
\section{Approach} \label{section:approach}
 
\begin{figure*}[htb]
	\centering
	\includegraphics[width=1\textwidth]{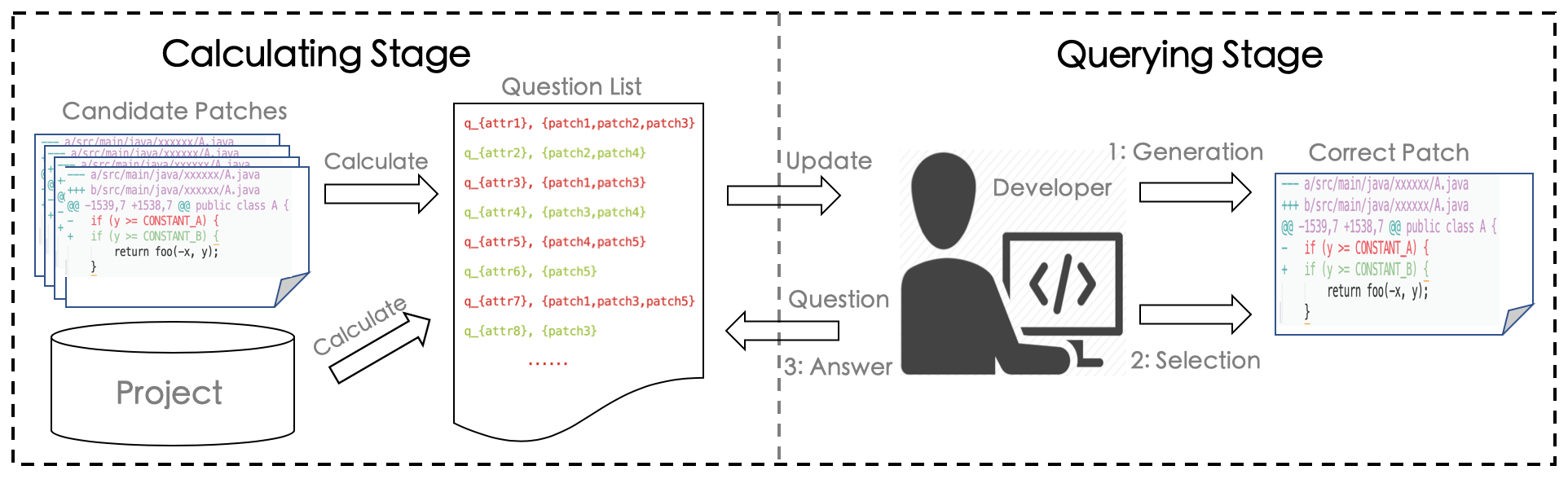}
	\caption{The workflow overview of the proposed approach.}
	\label{figure:workflow}
\end{figure*}

 
 In this section, we will first use a running example to introduce the overview of our approach in \secref{subsection:Overview}. Then, the next two sections (i.e., \secref{subsection:Calculating} and~\ref{subsection:Querying}) will demonstrate the two stages of our approach. \figref{figure:workflow} presents the overview of our approach.

\subsection{Overview}\label{subsection:Overview}
%

In this section, we will introduce our approach with a running example. \figref{fig:codesnippet} shows a code snippet from the buggy program Math41 in Defects4J~\cite{just2014defects4j} benchmark, which invokes the buggy method \code{evaluate()} when the condition \code{length>1} is satisfied. Besides, the following three patches are generated by existing APR techniques and can make all the test cases pass. In particular, $p_1$ and $p_2$ are incorrect patches that change the condition in line 320 (in \figref{fig:codesnippet}) to a new one, while $p_3$ is the correct patch that updates a \code{for} statement in the buggy method \code{evaluate()}.

\begin{figure}[htb]
\begin{lstlisting}[style=java,numbers=none]
<@{\tiny 313}@>   public double eval(double values, ...) {
       <@$\cdots$@>
<@{\tiny 320}@>     if(length == 1){  //incorrect patches change here
<@{\tiny 321}@>        var = 0.0;
<@{\tiny 322}@>     }else if(length > 1){ ...
<@{\tiny 323}@>        // buggy code resides in method evaluate()
<@{\tiny 324}@>        var=evaluate(values,weights,m,begin,length);
<@{\tiny 325}@>     }
       <@$\cdots$@>
<@{\tiny 329}@>   }
\end{lstlisting}
\caption{\label{fig:codesnippet} A code snippet from Math41}
\end{figure}

\newcommand{\codeRev}[2]{{\footnotesize {\textcolor{black}{\code{#1}}}$\rightarrow$ {\textcolor{blue}{\code{#2}}}}}
\begin{itemize}[leftmargin=30pt]
	\item [$\times$] $p_1$: \codeRev{if(length==1)}{if(length==5\&\&length!=0)}
	\item [$\times$] $p_2$: \codeRev{if(length==1)}{if((length\&1)==1)}
	\item [$\checkmark$] $p_3$: \codeRev{for(i=0;i<weights.length;)}{for(i=begin;i<begin+length;)}
\end{itemize}

By analyzing the program and the patches, we can collect a set of program attributes related to each patch. For example, when applying the first two patches (i.e., $p_1$ and $p_2$), the failed test case executes the statement in line 321. On the contrary, when applying patch $p_3$, the statement in line 321 is not executed. Therefore, checking the correctness of the attributes can help to filter incorrect patches. As a consequence, our approach will select the attributes that have the ability to distinguish different candidate patches, and treat them as questions to ask for developers' confirmation.
For example, for the above three candidate patches, the first two questions listed below correspond to the attributes related to program execution trace and change location, respectively.

\begin{itemize}
	\item []$q_1$: {\it Whether the statement in line 321 should be covered?}
	\item []$q_2$: {\it Whether the method \code{evaluate()} should be patched?}
	\item []$q_3$: ...
\end{itemize}

Then, we can present all the questions to developers. The developer could pick a question and provide an answer. For example, suppose the developer first select the first question, and regards it as incorrect (i.e., the answer is No.). The patches related to this attribute can be filtered (i.e., $p_1$ and $p_2$ in the example), because the patches cause the incorrect program attributes. However, the other patch (i.e., $p_3$) will be remained since it does not make the program have the attribute. In fact, each time a question (not the last one) is answered, there must be some patches filtered: when an attribute is refuted, the corresponding patches can be filtered. Otherwise, the other patches can be filtered. For example, the answer to question $q_2$ is confirmed, the patches $p_1$ and $p_2$ can be filtered, because they change the code in method \code{eval()} but not \code{evaluate()}. 

In summary, by answering questions, the number of candidate patches will monotonously decrease. Based on this insight, we proposed our approach called \toolname, which leverages program attributes as questions to ask developers and filters patches generated by APR techniques according to the answers. 

There is a challenge in implementing \toolname: since collecting some kinds of attributes may take a lot of time, such as program execution trace, it will be impractical to provide a timely response for online debugging. To overcome this challenge, we utilize the fact that repair approaches are assumed to work offline, e.g., after a daily build and before the working time of the next day. In fact, current repair techniques often require hours to fix a bug, and cannot be used online. Based on this fact, we design a two-stage approach. The first stage (i.e., Preparing Stage) is an offline process that collects program attributes, while the second stage (i.e., Interactive Stage) is an online process that only needs to filter patches based on the collected attributes, which can be achieved within a short response time.

%
%

\subsection{Preparing Stage}\label{subsection:Calculating}

In this section, we will describe the first stage in our approach, which is called preparing stage. As explained, this stage is an offline process that performs data preparation for the next interactive stage. In particular, when given a set of patches related to a bug, our approach automatically collects program attributes for different patches, which will be finally leveraged to construct a set of questions for interaction. 

Generally, many kinds of attributes can be used in our approach as long as they can distinguish the candidate patches from some perspective. However, the attributes that can distinguish more patches and are easy to understand for developers should be preferred, because they potentially can decrease the number of interactions and reduce the burden of developers. As a result, the current implementation of our approach employs three kinds of attributes of programs, which include both static code property (\method) and dynamic runtime features (\trace and \variable). The followings describe the details of the attributes.

\begin{itemize}
	\item {\bf \method} denotes the specific method in the program, which the patches modify the code in. This kind of attribute is described as ``\textit{The method m should be patched}'', where \textit{m} represents the name of some method. Therefore, according to the given patches, our approach automatically analyzes the change locations for each patch, i.e., which methods are modified by the patch. Particularly, when a patch changes multiple methods, our approach will record all of them.
	
	\item {\bf \trace} means the executed statements while running the failing test case over the patched program. For simplicity, we only consider execution traces in methods that are modified by the given patches. That is we first collect all methods that are modified by at least one patch, and then record the traces of the test execution in all those methods over the patched program. In particular, the traces are collected at the line level, i.e., which lines of code are executed.
	Additionally, since the same method may also be executed multiple times in one execution, we leverage a hierarchical aligning algorithm to compare the difference of execution traces. That is when given two traces, we first align them at method level and obtain pair-wise methods, and next we compare the traces in a pair of methods. In these two processes, we greedily align the traces based on the execution order. Finally, we identify the differences between traces at line level as attributes, i.e., some lines of code are uniquely executed over a part of patched programs. In general, this attribute is described as ``\textit{The statement at line n in method m should be executed}'', where \textit{n} and \textit{m} represent the line number and method, respectively. Therefore, each attribute corresponds to a unique line of code in the program.
	 
	
	\item {\bf \variable} indicates a variable is assigned some value at specific locations during the execution. Particularly, we consider all local variables and class fields with primitive types at the entry and exit locations of the modified methods. More concretely, for each method that is modified by at least one patch, we collect all values assigned to variables at the entry and exit of all invocations to the method. Therefore, the attribute of ``\textit{the value val assigned to var}'' denotes the variable \textit{var} is assigned the value of \textit{val} in the execution at least once over the patched program.
	
\end{itemize}


Therefore, in the current implementation of \toolname, in total we use three kinds of attributes. They are described as (1) the method \textit{m} should be patched, (2) the statement at line \textit{n} in method \textit{m} should be executed, and (3) the value \textit{val} assigned to \textit{var} is correct, respectively, where the \textit{m}, \textit{n}, \textit{val} and \textit{var} correspond to some method, line number, variable value and variable name.

In order to store the attributes and corresponding patches, we treat each attribute as a question and define the following data structure.

\begin{definition}
	({\it Interactive Question (IQ).}) An interactive question is a pair $\left\langle q_{attr}, patches \right\rangle$, where $q_{attr}$ is a question about whether the attribute $attr$ of the program is correct or not, and $patches$ is a set of patches that make the answer to question $q_{attr}$ {\it yes}, i.e, after applying the patch in $patches$, the program attribute $attr$ holds.
\end{definition}

For instance, one interactive question for the example presented in \secref{subsection:Overview} can be $\left\langle q_{attr1}, \{p_1, p_2\} \right\rangle$, where, $q_{attr1}$ is ``\textit{Whether the statement at line 321 in method \code{eval()} should be covered?}''. In this way, for each attribute, we can construct an interactive question, which will be used on the next interactive stage. In particular, we will delete {\it IQ}s whose $patches$ includes all candidate patches ahead of time to reduce the number of questions.

\subsection{Interactive Stage}\label{subsection:Querying}

As it is introduced above, the interactive stage is an online process with developers (shown in \figref{figure:workflow}) using the interactive questions constructed in the preparing stage. The input of this stage is a list of {\it IQ}s and the complete project under debugging. Each time, our approach collects the feedback from developers for some questions and update the candidate questions and patches in accordance. More concretely, in the interactive debugging process, there are in total three kinds of actions that a developer can take.

\newcommand{\answer}{{\bf {\em Answer}}\xspace}
\newcommand{\select}{{\bf {\em Selection}}\xspace}
\newcommand{\generate}{{\bf {\em Generation}}\xspace}

\begin{itemize}
	\item  \answer Answering an {\it IQ} to filter out some plausible but incorrect patches.
	\item \select Selecting a patch from the candidates in {\it IQ}s as the correct patch.
	\item  \generate Generating a correct patch by themselves to fix the bug.
\end{itemize}

The \answer action is the main procedure in the interactive querying process, which interactively refine the candidate patches with removing incorrect ones according to the answers of some question from developers.
Algorithm~\ref{algorithm:update} demonstrates the updating process for each round of interaction after developers answering a question. In this process, we maintain a list of {\it IQ}s (i.e., $\mathbb{Q}$) waiting for answers and a set of candidate patches (i.e., $\mathbb{P}$). In the beginning, $\mathbb{P}$ contains all candidate patches. In each round, we update these two parts according to the developer's answer. Particularly, when an attribute is correct (i.e., the answer is {\it yes} to the corresponding question), the patches related to it will form the new candidate patch set (lines 3-4), otherwise, they will be filtered from candidates (line 6). Finally, the candidate questions $\mathbb{Q}$ will be updated according to remaining patches (lines 8-13). In particular, if all patches are filtered in an {\it IQ}, it will be deleted and not require developers to answer in the future (line 10).

Additionally, in the debugging process, developers may also check the correctness of candidate patches when $\mathbb{P}$ is not too large. Therefore, the other two actions denote that developers can {\it select} the correct ones directly from candidate patches or {\it generate} patches manually and finalize the debugging process.

\begin{algorithm}  
	\caption{Update Algorithm}
	\label{algorithm:update}
	\begin{algorithmic}[1] 
		\Require ~$\mathbb{Q}$: question list, $\mathbb{P}$: all candidate patches 
			\Statex \indent\indent $q$: an answered question, $a$: answer to $q$
		\Ensure $\mathbb{Q'}$: updated question list, $\mathbb{P'}$: updated patch list 
		\State $\mathbb{Q'} \gets \emptyset$,  $\mathbb{P'} \gets \emptyset $
		\If{$\mathbb{Q} != \emptyset$ \&\& $\mathbb{P} != \emptyset $}
		\If{$a$ == yes}  \COMMENT{$q.attr$ is correct}
			\State $\mathbb{P'} \gets q.patches$ \COMMENT{patches satisfy $q.attr$}
		\Else
		    \State $\mathbb{P'} \gets \mathbb{P} \setminus q.patches$ \COMMENT{patches do not satisfy $q.attr$}
		\EndIf
		
		\For {each $q' \in  \mathbb{Q}$}		  \COMMENT{update candidate questions}
			\State $q'.patches \gets q'.patches \cap \mathbb{P'}$
			\If{$q'.patches \neq \emptyset$ }  
			\State $\mathbb{Q'} \gets \mathbb{Q'} \cup  \{ q'\} $
			\EndIf
		\EndFor
		\EndIf

	\end{algorithmic}  
\end{algorithm}

\section{Eclipse Plugin} \label{section:implementation}

\newcommand{\qv}{{\em Query View}\xspace}
\newcommand{\dv}{{\em Diff View}\xspace}

\begin{figure*}[htbp]
	\centering
	\includegraphics[width=1 \textwidth]{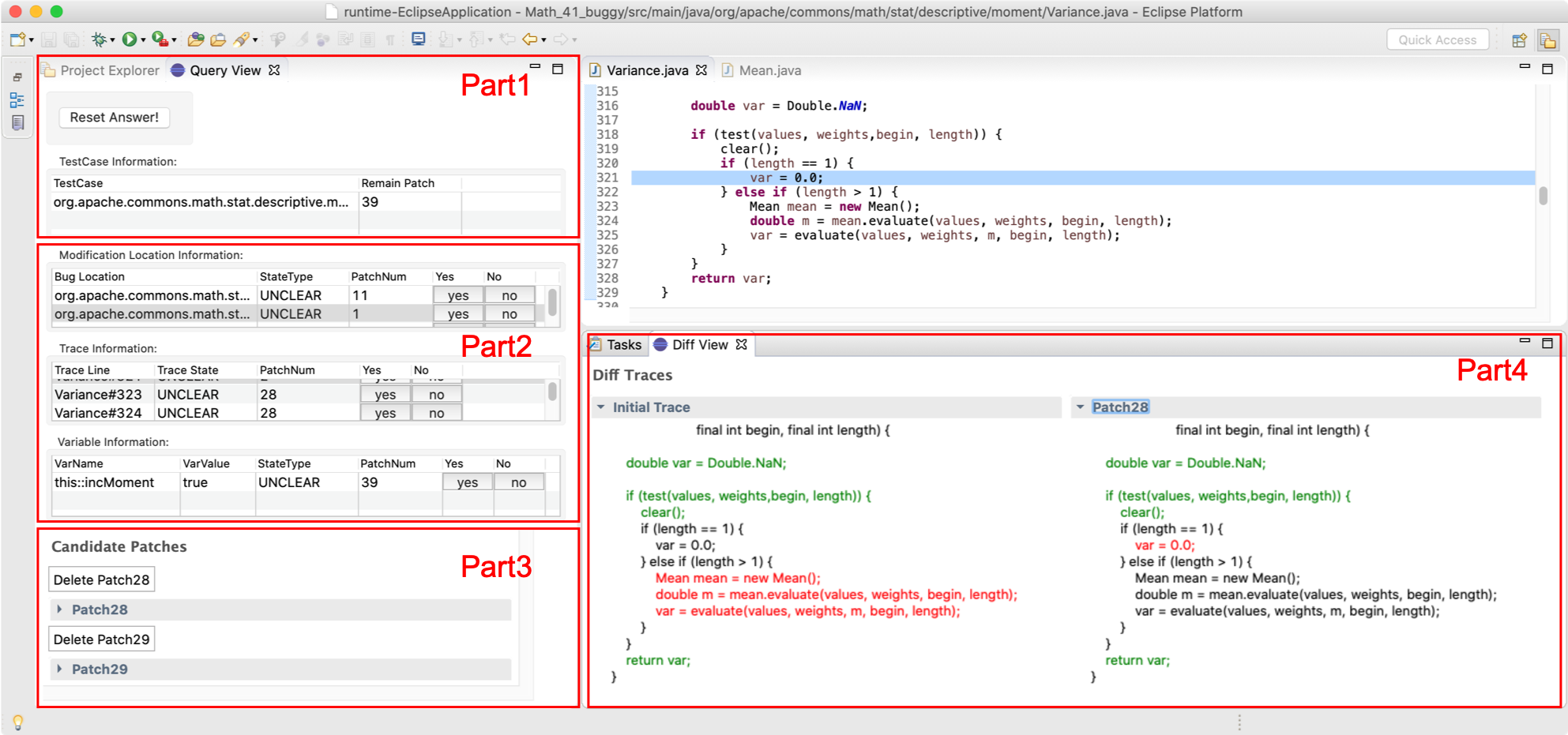}
	\caption{The screenshot of \toolname.}
	\label{figure:gui}
\end{figure*}

To evaluate the effectiveness of our approach, we have developed a prototype tool called \toolname, which is a plugin program for Eclipse with a graphical user interface (GUI). \figref{figure:gui} shows a snapshot of the plugin during a debugging process. Specifically, it consists of two embedded views, \qv and \dv, for collecting developers' feedback and displaying information to developers.

\qv is the main component of our approach, which presents the details of {\it interactive question}s and corresponding candidate patches. In order to separately display different kinds of information, it is further subdivided into three panels. As shown in the figure, the first panel shows failing test cases and the number of candidate patches thus far. The second panel shows the details of questions that developers can selectively answer. As introduced in \secref{subsection:Calculating}, currently we employed three kinds of attributes that corresponds to the three groups of questions in the view. For each question, it not only displays the attribute details, but also shows the number of related patches and the state the of question. When a question is answered ({\em Yes} or {\em No}), the state of the question will be updated from UNCLEAR to YES/NO. As a separate panel, we display the candidate patches when a question is selected. Additionally, the plugin also provides a one-click rollback to reset all the answers. 

\dv is an auxiliary view to visualize the differences of execution traces before and after applying a patch to the buggy program, where the \textcolor{green}{green} lines of code are commonly covered by the failing test before and after repair, while the \textcolor{red}{red} lines of code are particularly covered by one of them. Finally, the other lines of code are changed by the patch or not covered by any of them. In this way, the developers can clearly understand the impact of the patch on the program execution, and possibly feel easy to answer the questions.

Please note that all views or panels are logically interrelated to each other, the selection of one part may trigger the update of the display in other places. For example, when a patch is selected, the \dv will refresh the trace difference immediately. Moreover, developers can locate the changed code in the editor by simply selecting a patch. Most importantly, when a question is answered, the candidate patches of all other questions will be updated according to Algorithm~\ref{algorithm:update} and refreshed on the view.

\section{Evaluation} \label{section:evaluation}
To evaluate the effectiveness of \toolname, we have conducted two experiments. The first one is a simulation experiment, which investigates the effectiveness and efficiency of \toolname when applied to a large number of real-world bugs. Besides, in this experiment, we also study the impacts of different kinds of {\it interactive questions}. The other experiment is a user study to evaluate the usefulness of \toolname in a realistic program repair scenario. Specifically, we investigate whether it can improve the efficiency and correctness of developers when debugging.

\subsection{Simulation Experiment}
In this experiment, we designed the study to answer the following research questions:
\begin{itemize} 
	\item \textbf{RQ1}: How effective is \toolname in debugging real-world bugs?
	\item \textbf{RQ2}: How effective are different kinds of {\it interactive questions} of \toolname?
\end{itemize}

RQ1 investigates the effectiveness and efficiency  
of \toolname via the number of remaining patches and queries. Ideally, all incorrect patches can be correctly filtered out and only correct patches (can be empty) are left after several rounds of queries.
RQ2 compares the effectiveness of questions built with different kinds of constraints.




\subsubsection{Experiment Setup}
\paragraph{Dataset}
\label{subsection:dateset}
In order to simulate the scenario, where there are multiple patches for a bug, we can apply patches produced by multiple existing APR techniques. We consider all existing automatic program tools, which work on Java language and are available thus far. As a matter of fact, different APR techniques may produce similar or even the same patches, to improve the efficiency and clarity for interactive debugging, we will remove duplicate patches ahead of time, which are the same in syntax. In total, we selected 13 program repair tools and the details of each tool are presented in \tabref{tab:aprtools}. All of them were evaluated on the commonly used Defects4J~\cite{just2014defects4j} benchmark and the results are available. We collected the patch data from previous studies~\cite{martinez:hal-01321615,kPAR-url,liu2019avatar,Wen2018ContextAwarePG,ISSTA18-SimFix,Xuan2016History,xiong-icse17,xuan2016nopol}. In total, we have collected 8654 patches for 85 bugs, and the details are listed in \tabref{tab:dataset}. In the table, the first two columns present the project names and line numbers of source code. Columns ``{\bf Bug}'' and ``{\bf AvgPatch}'' show the number of bugs in each project and the average number of patches for each bug. Finally, the column ``{\bf C/NC}'' denotes the number of bugs that have or do not have correct patches collected.


\begin{table}[htp]
	\begin{adjustbox}{width=0.45\textwidth,center}
		\begin{threeparttable}
			\small
			\centering
			\caption{Dataset in experiments.}
			\label{tab:dataset}
			\begin{tabular}{l|cccc}
				\toprule
				{\bf Project} & {\bf kLoC} & {\bf Bug} & {\bf AvgPatch} & {\bf C/NC} \\
				\midrule		
				JFree\textbf{Chart} & 96 & 17 & 225 &  9/8 \\
				\textbf{Closure} Compiler & 90 & 13&  6 & 3/10 \\
				Apache Commons \textbf{Lang} & 22 &13 & 66 &  8/5 \\
				Apache Commons \textbf{Math} & 85 & 42 & 92 & 15/27 \\
				\midrule
				{\bf Total} & 321 & 85 & 101 & 35/50\\
				\bottomrule
			\end{tabular} 
		\end{threeparttable}
	\end{adjustbox}
\end{table}

\begin{table}[htb]\small
	\centering
	\caption{APR tools included by \toolname.}
	\label{tab:aprtools}
	\begin{tabular}{l|m{6cm}}
		\toprule
		{\bf Name} & {\bf Description} \\
		\midrule		
		jKali & The Java implementation of Kali~\cite{PatchPlausibility}, which only performs functionality deletion. \\
		\hline
		jGenProg & The Java implementation of GenProg~\cite{GenProg,GenProgTSE}, which repair bugs with genetic programming algorithm. \\
		\hline
		kPAR & The Java implementation of PAR~\cite{PAR}, which generate patches based on predefined fix patterns.\\
		\hline
		Nopol~\cite{xuan2016nopol}  & Relying on constraint solving to fix incorrect conditions. \\ 
		\hline
		jMutRepair~\cite{martinez:hal-01321615} & A mutation based program repair tool.\\
		\hline
		Cardumen~\cite{DBLP:conf/ssbse/MartinezM18} & Generating patches based on mined templates.\\
		\hline
		Avatar~\cite{liu2019avatar} & A repair tool based on the fix patterns of static analysis violations.\\
		\hline
		HDrepair~\cite{Xuan2016History} & A repair tool based on historical bug-fix information.\\
		\hline
		ACS~\cite{xiong-icse17} & Learning statistical information from open source programs for fixing incorrect conditions.\\
		\hline
		3sfix~\cite{ChenThe} &  \multirow{3}{*}{Repair approaches based on similar code match.} \\
		CapGen~\cite{Wen2018ContextAwarePG} & \\
		SimFix~\cite{ISSTA18-SimFix} & \\
		\hline
		DeepRepair~\cite{White2017} & An extension of jGenProg, which leverages code similarity.\\
		\bottomrule
	\end{tabular} 
\end{table}

\paragraph{Procedure}
To automatically simulate the interaction process with developers, each time \toolname randomly selects one question from all candidates and then automatically gets the answer via analyzing the fixed programs. Moreover, to alleviate the impact of randomness, we repeat the interaction process for each bug five times and take the mean number of queries as the final result.

\subsubsection{Results for Remaining Patches and Query Number (RQ1)}
Table~\ref{tab:results_rq1} shows the results of this research question. In the table, columns ``{\bf None}'' and ``{\bf All Correct}'' show the number of bugs, which do not have any candidate patch left and only have correct patches left, respectively. The following columns present the percentage ranges of patches left among all candidates. For example, ``$\leq 40\%$'' denotes that the percentage is from 20\% to 40\%. Each cell shows the number of corresponding bugs. Particularly, we separately display the number of bugs that contain (\textbf{Con Bug}) and do not contain (\textbf{NotCon Bug}) correct patches. Finally, row ``\textbf{Query Number}'' shows the average number of queries.

From the table, our approach can correctly filter out all incorrect patches and only make the correct ones left for 49.4\% (42/85) bugs in total. Particularly, when there is no correct patch, it can filter all incorrect patches for about 78\% (29/50) bugs. Additionally, after analyzing these bugs we find that the number of candidate patches ranges from 2 to 1248, and on average more than 60 per bug, which may potentially cost a lot of time of developers to manually review. However, in this process, our approach on average only requires about 3.2 queries for each bug after filtering all of them. As explained in our user study (\secref{subsec:userstudy}), the interactive debugging process will not cause a big burden to human developers and significantly improve the efficiency of manual review. On the other hand, when the correct patches are given, our approach can help correctly remove all incorrect patches while still save the correct ones for about 37.1\% bugs (13/35), and the number of queries is even smaller, i.e., three queries on average. The result indicates that our approach is effective for patch filtering. 

\finding{For about 49\% bugs, \toolname can filter out all incorrect patches and save all correct patches within 3.1 questions on average. }

However, from the table we can see that there are still some incorrect patches that cannot be completely filtered out in the experiment. For example, there are 3.5\% (3) bugs having less than 20\% (or 20\%-40\%) candidate patches left. However, the queries needed are still not too many, usually less than six per bug apart from the one which needs 20 queries (in $\leq$ 20\%). The main reason is the candidate patches are semantically too similar to each other, causing the program executions the same. For example, \figref{fig:chart1} shows two candidate patches, where the first one (left) is the correct patch while the other is incorrect. However, both these two patches change the \code{if} condition (line 679) and have the same execution path, making our approach cannot better distinguish them. However, more attributes can be added to further improve the effectiveness of our approach. Additionally, even though some patches cannot be filtered out by our approach, they are possibly easy for developers to review (e.g., the incorrect patch in \figref{fig:chart1} compares two constant values). As we will explain in the user study (\secref{subsec:userstudy}), even if 12/26 (>40\%) candidate patches left in Task1, our approach can still improve the efficiency of manual review process.


\begin{figure}[htb]
\begin{lstlisting}[style=java,numbers=none]
//correct patch         //incorrect patch
<@{\tiny 679}@>- if(dataset!=null){  <@{\tiny 679}@>- if(dataset!=null){
<@{\tiny 679}@>+ if(dataset==null){  <@{\tiny 679}@>+ if(AbsRenderer.ZERO==null){
<@{\tiny 680}@>     return result;   <@{\tiny 680}@>	  return result;
<@{\tiny 681}@>  }                   <@{\tiny 681}@>  }
\end{lstlisting}
\caption{\label{fig:chart1} Example candidate patches left after filtering.}
\end{figure}


Finally, for about 23.5\% (20/85) bugs, \toolname cannot generate any questions to filter candidate patches. We further reviewed these bugs and found the reason is that most of the patches are modified the same location and have similar program attributes, which cannot be better distinguished by the current implementation of \toolname. In fact, 16 out of the 20 bugs contain less than 5 candidate patches. When the number of patches is small, it may be easy for the developer to review.

\begin{table}
	\begin{adjustbox}{width=1.0\columnwidth,center}
	\begin{threeparttable}
		\small
		\centering
		\caption{Remaining patches in bugs.}
		\label{tab:results_rq1}
		\begin{tabular}{l|cc|cccc|c}
			\toprule
			Remain Patches & None &  All Correct & $\leq 20\% $& $\leq 40\% $ & $ >40\%$ & $ =100\%$ & Total \\
			\midrule
			Con Bug & 0 & 13 & 2 & 1 & 13  &  6 & 35\\
			NotCon Bug & 29 & 0 & 1&  2 & 4  &  14 & 50 \\
			\midrule
			Total Bug & 29  & 13 &  3 &  3 & 17 & 20 & 85 \\
			\midrule
			Query Number & 3.2 & 3.0 & 9.8 & 1.7 & 2.2 & -- & 3.1 \\ 
			\bottomrule
		\end{tabular}
	\end{threeparttable}
\end{adjustbox}
\end{table}

\subsubsection{Results for Different Questions (RQ2)}
To investigate the effectiveness of different attributes in our approach, we conduct a controlled experiment, where each time we apply questions based on only one kind of attribute. In addition, as explained above that the current attributes do not have sufficient ability to distinguish all candidate patches. In the comparison, we only focus on the patches which can be filtered by \toolname.

\begin{figure}
	\begin{adjustbox}{width=0.35\textwidth}
		\centering
		\includegraphics[width=1 \textwidth]{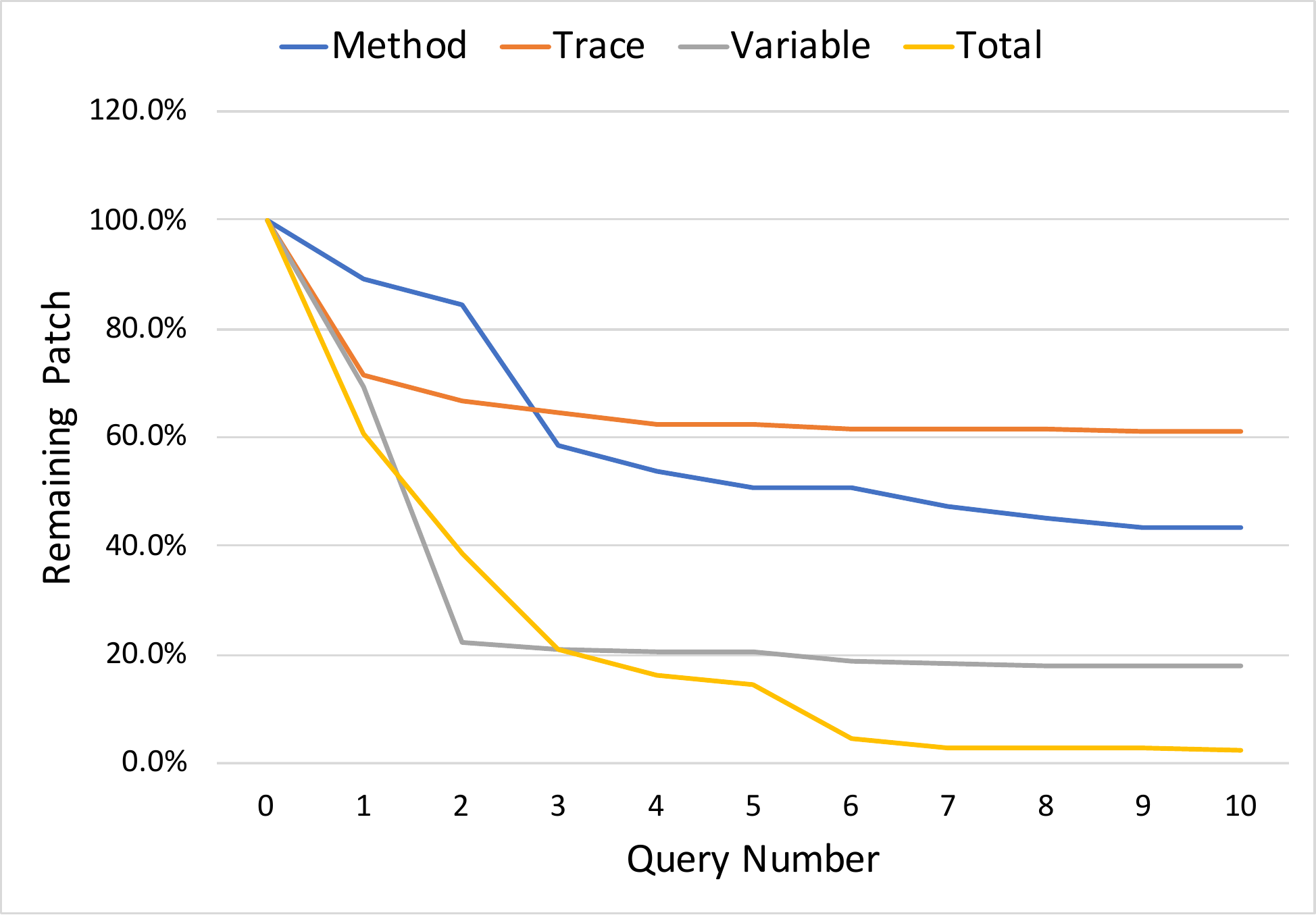}
	\end{adjustbox}
	\caption{Remaining patches and query numbers.}
	\label{fig:remainPatch}
\end{figure}

\figref{fig:remainPatch} shows the experimental results when applying different attributes. 
The $x$-axis denotes the number of queries while the $y$-axis denotes the percentage of remaining patches. 
From the figure, the performances of different kinds of attributes vary greatly, and there is no such attribute that can filter all incorrect patches. 
The reason is incorrect patches tend to be similar to each other on a certain attribute while different on some other attributes.
For example, the attribute of \variable can at most distinguish about 82\% incorrect patches. 
After combining other two attributes, it can filter up to  97\% incorrect patches in ten queries.


In addition, the performance when using \variable is even better than that using all attributes within three queries. This is because of the randomness in the simulation process. However, though \variable is effective, developers seldom use it in practice based on the developers’ feedback, since it is usually hard to answer. On the contrary, the questions related to the attribute of \method is the most frequently selected as it is easy to answer and performs relatively well in all cases.

Nevertheless, the results show that the number of queries is usually no larger than 6 to achieve the best performance for each attribute. Besides, usually the first two queries contribute most.

\finding{ The performance of \variable is better than the performance of other two attributes, but it is worse than that of combining all attributes. }

\subsection{User Study}
\label{subsec:userstudy}
Since \toolname was designed as an interactive debugging approach, to investigate whether it helps developers in practice, we further conducted a user-in-the-loop study, which focused on the following research questions:
\begin{itemize} 
	\item \textbf{RQ3}: How do developers perform while debugging with \toolname?
	\item \textbf{RQ4}: What is the feedback of developers after using \toolname?
\end{itemize}

To investigate the impacts of our approach to developers' debugging performance in practice, we configured three debugging settings: \manualFix, \fixwithPathes and \fixwithtool. \manualFix and \fixwithPathes denote that the developers manually repair bugs without and with patches, which are produced by the 13 APR tools used in our paper, respectively, while \fixwithtool denotes that the developers repair bugs with the assistance of \toolname. 
Additionally, after the experiment, we have interviewed all developers and collected their feedbacks for further analysis. 

\subsubsection{Study Design} 

We will introduce our study design from three aspects: \textit{Tasks}, \textit{Participants} and \textit{Procedure}.

\textit{Tasks.} We selected four bugs as debugging tasks that are not straightforward to repair. Additionally, we also considered the diversity of programs, i.e. from different projects, which perform different types of tasks. Table~\ref{tab:tasks} shows the task details, including the bug id and patch number for each bug, etc. All bugs are from the Defects4J dataset shown in \tabref{tab:dataset}. Particularly, Task1 and Task2 contain the correct patches for the bugs while Task3 and Task 4 do not. Besides, the participants would not be informed whether the candidate patch set contains a correct patch.

\begin{table}[htb]
	\begin{adjustbox}{width=1.0\columnwidth,center}
		\begin{threeparttable}
			\small
			\centering
			\caption{Tasks in user study.}
			\label{tab:tasks}
			\begin{tabular}{c|ccccc}
				\toprule
				Task ID & Bug ID & Patch & Correct Patch &  Query Number &  Remain Patch \\
				\midrule
				Task1 & Chart9 &  26 & 3 & 3 & 12\\
				Task2 & Math41 & 48  & 1 & 6 & 1 \\
				\midrule
				Task3 & Lang14 & 8 & 0 & 2.4 & 0 \\
				Task4 & Lang22 & 24 & 0 & 2.4 & 0 \\
				\bottomrule
			\end{tabular} 
			\begin{tablenotes}
				\item \emph{Note}: In this table, column \textbf{``Patch"} and column \textbf{``Correct Patch"} denote the number of original patches and the number of correct patches in original patches, respectively,
				and column \textbf{``Remain Patch"} denotes the number of remaining patch after all questions are answered.
			\end{tablenotes}
		\end{threeparttable}
	\end{adjustbox}
\end{table}

\textit{Participants.} In total, we recruited 30 participants to conduct our user study. They are all students who majored in computer science from our department. Besides, they have at least three-year programming experience and are familiar with debugging in Eclipse. Additionally, the participants have no prior experience of repairing those bugs in the study.



\textit{Procedure.} In the study, participants were evenly divided into three separate groups (i.e., A, B and C) with each group including 10 participants. Each participant would finish four tasks in the corresponding group as shown in Table~\ref{tab:group}. For example, the participants in Group A should manually repair the bugs in Task1 and Task3 and manually repair bugs in Task2 and Task4 with the help of patches. As a result, from the table, each participant would finish all four tasks under two different debugging scenarios. Therefore, our study consists of 120 ($30\times 4$) individual debugging processes and each debugging process is called one {\it debugging session}.

\begin{table}[htb]
	\begin{adjustbox}{width=0.45\textwidth,center}
		\begin{threeparttable}
			\small
			\centering
			\caption{Groups in user study.}
			\label{tab:group}
			\begin{tabular}{c|ccc}
				\toprule
				Group & ManuallyFix & FixWithPatches & FixWith\toolname \\
				\midrule
				Group A & Task1+3 &  Task2+4 & --\\
				Group B & Task2+4 & -- & Task1+3 \\
				Group C & -- & Task1+3 & Task2+4 \\
				\bottomrule
			\end{tabular} 
		\end{threeparttable}
	\end{adjustbox}
	
\end{table}

In addition, to make the participants familiar with our tool, before the formal user study, participants in Group B and C were required to debug an irrelevant bug using \toolname until they got familiar with \toolname. Since there are too many debugging sessions, we assigned each session 30 minutes. If the participants cannot finish the debugging within the given time slot, we considered the bug failed to be repaired. After the participants finished a session, we would manually check whether the patch, figured out by the developers, was right.

After finishing the debugging, we interviewed each participant and collected their feedbacks. In specific, for each group, we carried out the interview in terms of the question:
\textit{what was the difference between the two settings you have experienced?}
for the participants who have debugged with \toolname, we would ask more questions about the function of \toolname, such as, (\romannumeral1) \textit{Which kinds of attribute related questions were most useful in \toolname?} (\romannumeral2) \textit{Was Diff View useful for debugging?}

\subsubsection{Results for Repaired Bugs and Repair Time (RQ3)}
\label{subsubsec:rq4}

To measure the performance of developers, we consider both the number of debugging sessions where the bugs were correctly repaired and the time used during debugging. 

\tabref{tab:repairNumber} shows the number of sessions in which the bugs were successfully repaired in three different debugging scenarios. From the table, our approach (\fixwithtool) significantly outperformed  \manualFix and \fixwithPathes with respectively 62.5\% and 39.3\% improvements. Especially, when the correct patches existed in the candidate patches, developers could always fix the bug in the study. However, it was not the case for \fixwithPathes even though the same patches were given. The result demonstrated the effectiveness of our approach.

\begin{figure*}[htbp]
	\resizebox{\textwidth}{!}{
		\centering
		\subfigure[Task1 (Chart9)]{
			\includegraphics[width=5.5cm]{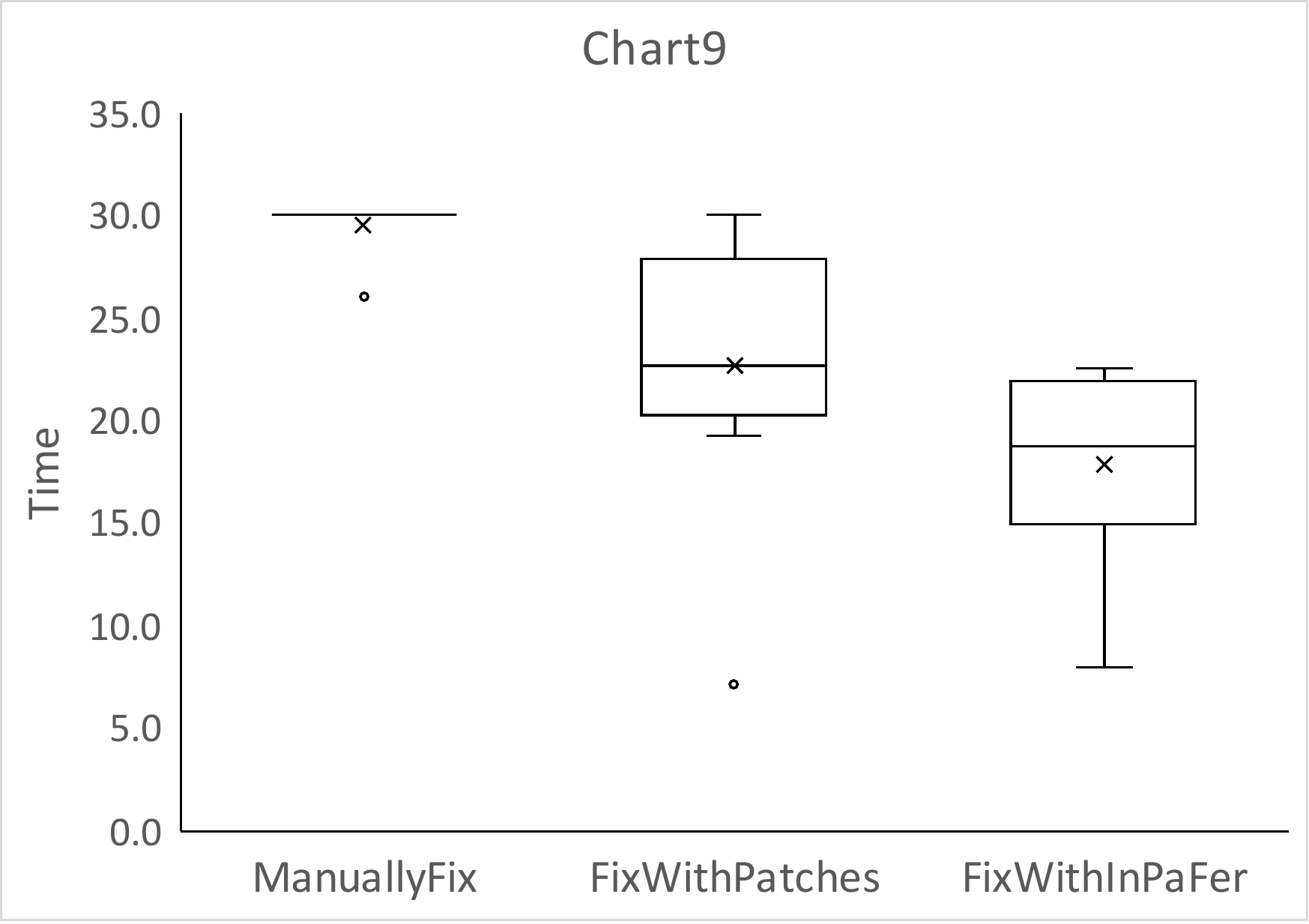}
			\label{subfig:a}
		}
		\quad
		\subfigure[Task2 (Math41)]{
			\includegraphics[width=5.5cm]{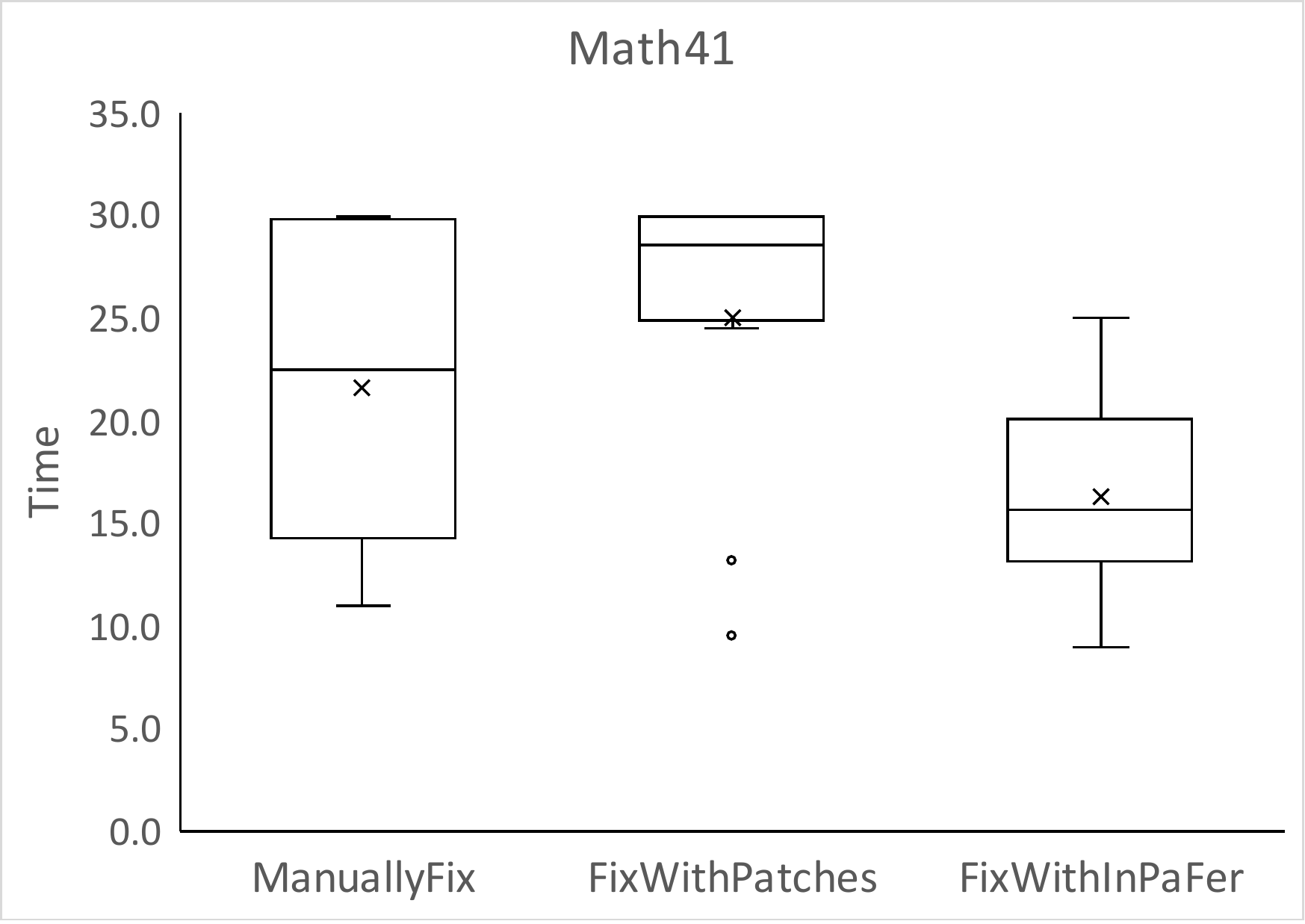}
			\label{subfig:b}
		}
		\quad
		\subfigure[Task3 (Lang14)]{
			\includegraphics[width=5.5cm]{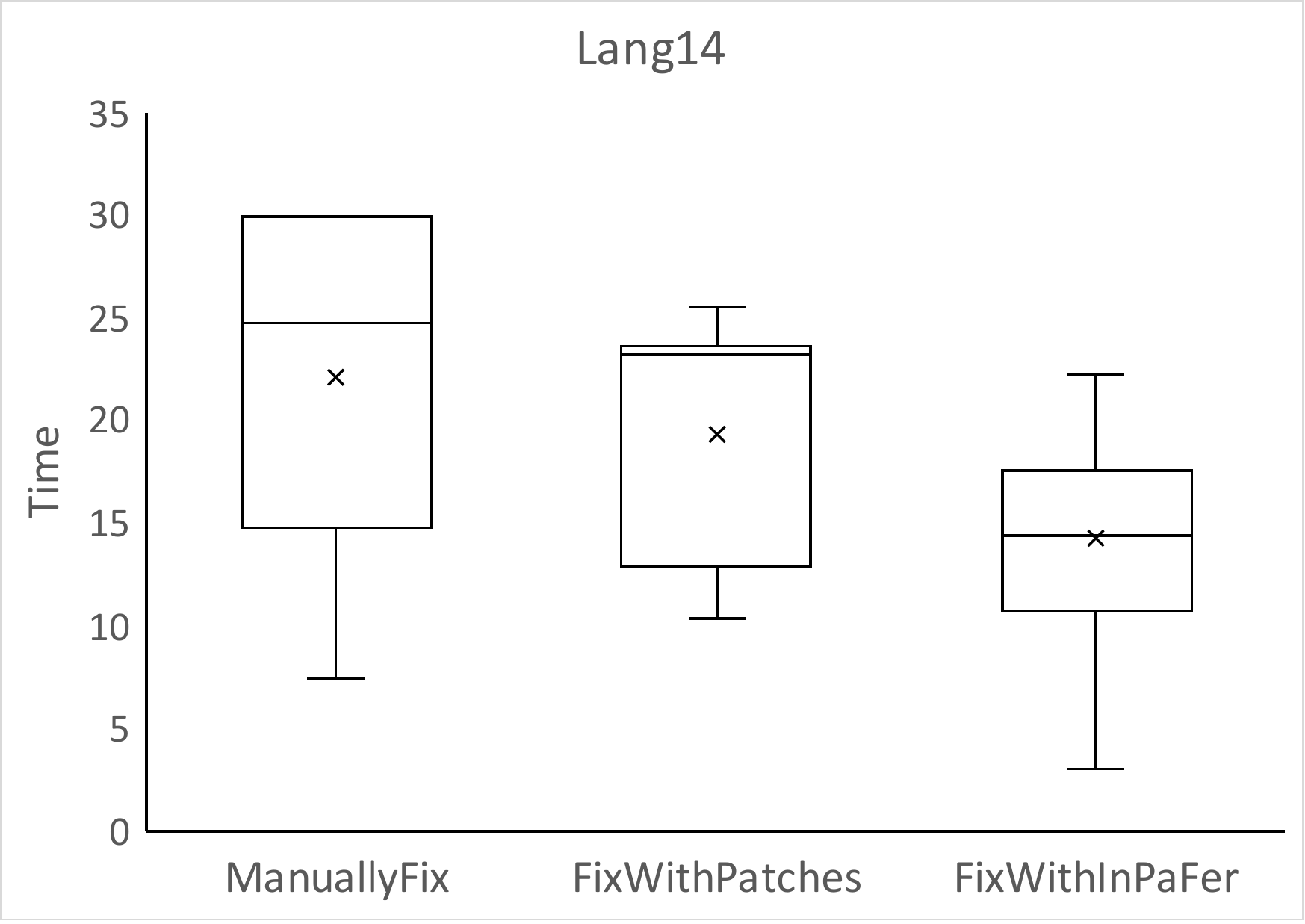}
			\label{subfig:c}
		}
		\quad
		\subfigure[Task4 (Lang22)]{
			\includegraphics[width=5.5cm]{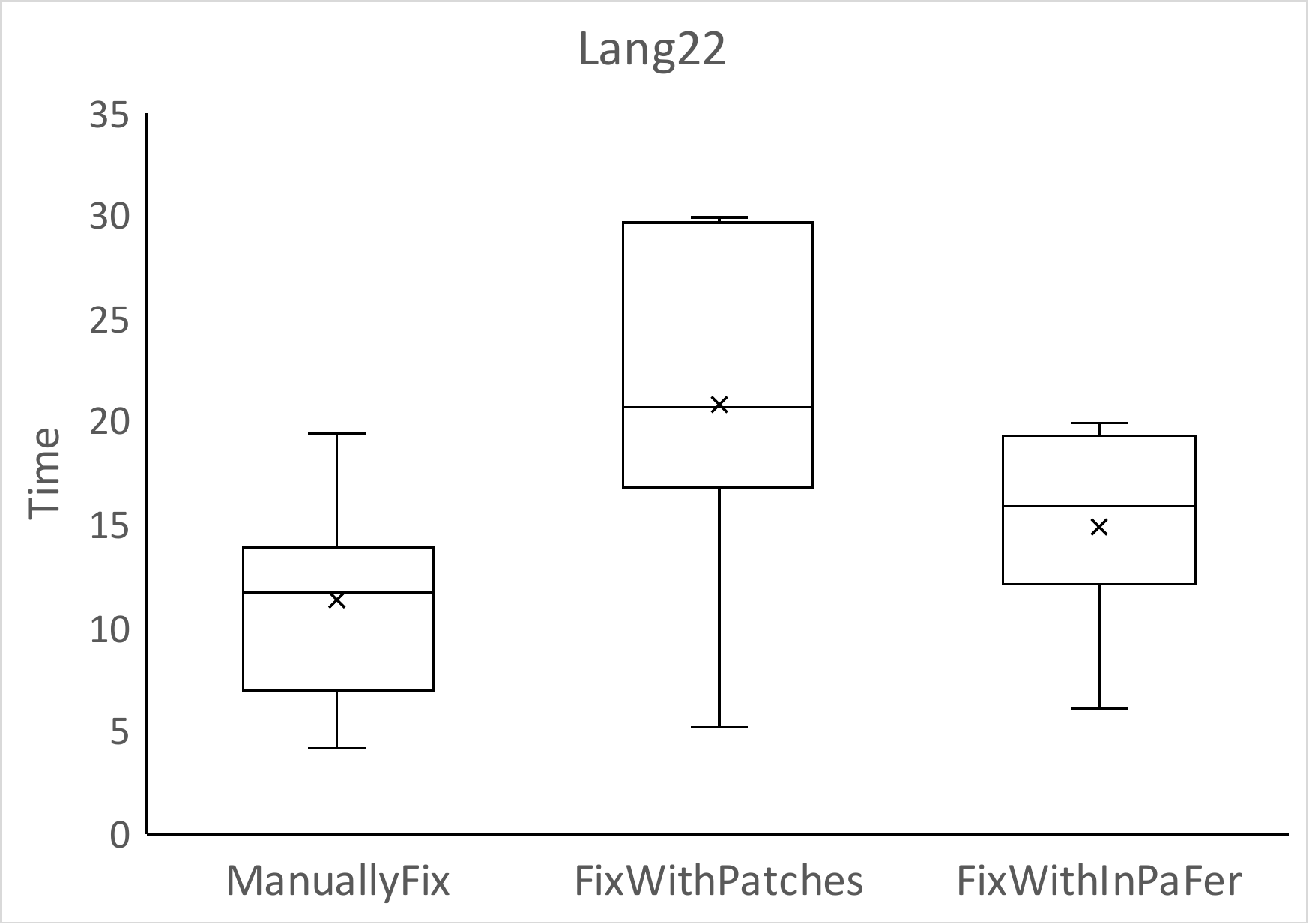}
			\label{subfig:d}
		}
	}
	\caption{Debugging time in user study.}\label{fig:userStudyTime}
\end{figure*}

Additionally, when considering the debugging time, our approach achieved better performance as well. Particularly, it could significantly shorten the debugging time of developers in all tasks compared with \fixwithPathes and in three out of four tasks compared with \manualFix. Overall, our approach could reduce the debugging time by 25.3\% and 28.0\% against the other two, respectively. Therefore, our approach can improve the efficiency of human debugging.

\finding{Overall, \fixwithtool can reduce the debugging time by 25.3\% and 28.0\% on average, and increase the success rate by 62.5\% and 39.3\% on average, compared to \manualFix and \fixwithPathes, respectively.}


\begin{table}[htb]
	\begin{adjustbox}{width=0.45\textwidth,center}
		\begin{threeparttable}
			\small
			\centering
			\caption{Number of successful debugging sessions in user study.}
			\label{tab:repairNumber}
			\begin{tabular}{c|ccc}
				\toprule
				{\bf Task ID} & {\bf \manualFix} & {\bf \fixwithPathes} & {\bf \fixwithtool}  \\
				\midrule
				Task1 & 1 & 9  &  10 \\
				Task2 & 8 & 6  & 10 \\
				\midrule
				Task3 & 5 & 8 & 10 \\
				Task4 & 10 & 5 & 9 \\
				\midrule
				{\bf Total} & {\bf 24} & {\bf 28} & {\bf 39} \\
				\bottomrule
			\end{tabular} 
		\end{threeparttable}
	\end{adjustbox}
\end{table}

From Table~\ref{tab:repairNumber} and Figure~\ref{fig:userStudyTime}, we observed that the relative performances of the developers in three debugging settings were different in four tasks. Specifically, in Task1 and Task3, \fixwithPathes had more successful debugging sessions but less debugging time compared with \manualFix, while it would reverse in Task2 and Task4. Therefore, we further investigated the reasons for this difference. Based on our data, we suspected that they were mainly due to the number and quality of candidate patches. More concretely, we make the following observations from the data:
\begin{enumerate}
	\item On the one hand, as shown in Table~\ref{tab:tasks}, Task1 contains 3 correct patches in 26 candidate patches, while Task2 contains only 1 correct in 48 patches. Although both Task3 and Task4 do not contain correct patches, Task3 contains only 8 incorrect patches, while Task4 contains 24 incorrect patches.
	This observation suggests that the number of incorrect patches is negatively related to the repair performance.
	\item On the other hand, we found that almost all candidate patches in Task1 exactly changed the faulty code, even if the patches are incorrect, and the candidate patches in Task3 provided partially correct code. The fault location and referable code potentially could provide guidance for developers to better understand the bugs. In contrast, the candidate patches in Task2 changed the code in different locations, and the candidate patches in Task4 provided meaningless code. These may mislead developers. In summary, different incorrect patches may have different quality and high-quality incorrect patches still guide the developers.	 
\end{enumerate}

\finding{High-quality incorrect patches with partial correct code at faulty location can still be helpful to developers.}

In addition, in Task1 and Task3, \fixwithtool shows small improvement than \fixwithPathes, while it shows greater improvement than \fixwithPathes in Task2 and Task4. This suggests that \toolname reduces the negative effect of the low-quality incorrect patches, and this reduction is more significant when the negative effect is larger.
 
\finding{ 
	The number and quality of incorrect patches affect the debugging performance of developers when they are provided with patches, and \toolname helps reduce the negative effect from low-quality incorrect patches.}
	
	
Besides, for Task1, Task2 and Task3, the performances of developers in \fixwithtool are significantly better than that in \manualFix, while the performance of developers in \fixwithtool is a little worse than that in \manualFix. 
This suggests that when containing correct or incorrect but high-quality patches, \toolname could significantly improve the performance of developers. On the other hand, according to the result of Task4, when providing patches with low quality, \toolname will not affect developers' performance too much and still can improve the manual patch review.

\subsubsection{Results for Feedback (RQ4)}
\newcommand{\rqstyle}[1]{\emph{#1}}
\newcommand{\keywd}[1]{\emph{#1}}
\newcommand{\ttle}[1]{\textbf{#1}}

To better understand the debugging process and the attitude of developers to our approach in practice, we conducted an interview with all participants after the experiments. The details are listed below.

\ttle{Group C: compare \fixwithtool and \fixwithPathes}

All participants of Group C gave a positive feedback on \toolname when repairing Task2. Based on the interview, we summarized the help from the following two aspects. 
(\romannumeral1) \keywd{Patch Filtering}. Most participants said that \toolname could help them to filter most of incorrect patches by answering a few questions, like \textit{``It can narrow the range of correct patches after a few simple questions."}. Therefore, they only needed to check several candidate patches.
(\romannumeral2) \keywd{Bug Understanding}. Since the developers were not familiar with the bugs they were debugging, it was hard to identify whether a given patch is correct or not. However, for Task2, the interactive questions were not hard to answer, and they provided a clue to help the developers to understand the bugs step by step. Finally, they could understand the bugs better and select the correct patches easier. 

Besides, for Task4, the participants thought that \toolname only help them to filter out all incorrect patches, but it cannot help them to repair the bugs, because the incorrect patches are meaningless code, which cannot provide other help.

\ttle{Group B: compare \fixwithtool and \manualFix}
 
The participants in Group B indicated that \toolname could provide the correct or partially correct patches, which provided guideline for debugging. Specifically, for Task1, they could find the correct patch after answering only a few questions. Besides, although Task3 does not contain the correct patch, the partially correct patches also help them a lot.  

The participants also referred to a limitation that \toolname does not support single answer cancellation. 
Specifically, the developers may misunderstand Task1 and select a wrong answer for some questions. This would lead to the correct patches wrongly filtered. When the developers realized that they selected a wrong answer, they need to reset all questions and restart the interaction again, which would waste the debugging time. 
This is a limitation of our current tool implementation but not our approach. A better tool implementation could make the answer to a question reversible, and potentially further boost the performance of \toolname.

 

\ttle{Group A: compare \fixwithPathes and \manualFix}

On the one hand, small part of participants thought that candidate patches could provide the \keywd{Fault Location} for Task4, which could provide guidance when debugging.
On the other hand, almost all participants complained that it was difficult to find the correct patch as there are too many patches to review when repairing Task2. Reviewing many incorrect patches would disturb their debugging.


\finding{ 
	Answering the questions of \toolname helps filter out incorrect patches as well as understanding the bug. }

\ttle{For all participants who fixed bugs with \toolname, we carried out the interview in terms of the following questions:}

\begin{enumerate}[leftmargin=0pt,itemindent=.75cm]		
	\item \rqstyle{Which kind of attributes related questions were most useful in \toolname?} Though we have already compared the impacts of different kinds of attributes related questions in the study, the result is still unclear to us from the developers' perspective. In the interview, almost all participants regarded that the questions related to \trace helped them most. Particularly, a developer explained that \textit{``developers know where the programs should execute, but they don't know what's wrong. \toolname knows all executed location, but it does not know whether the execution flow is correct or not. It is very helpful to combine these two kinds of information."} 
	Besides, a small number of participants also agreed that \method was also somehow helpful. They said that when the method was easy to understand, the question related to \method could be helpful, otherwise, it would be hard to answer this question as a much deeper understand of the bug was needed. However, they suggested that it would be more useful to developers who was familiar with the project. Finally, the questions related to \variable were the least to be selected as useful,
	because methods are often invoked multiple times during an execution, and the current question type does not allow us to locate a specific invocation.
	
	
	\item \rqstyle{Was Diff View useful for debugging?} About a third of participants thought that it was useful to understand the bug and correct the misunderstanding. For example, one participant said that, \textit{``The Diff View shows which branch that the execution get into makes the test case pass. It corrects the previous misunderstanding."}
\end{enumerate}

\finding{ In our user study, the questions related to \trace helped the developers most, while \variable could be improved by distinguishing different invocations.}

\section{Threats to Validity} \label{section:threats}
The \textit{internal} threat to validity lies in the recruited participants. On the one hand, all of them are not familiar with the projects, which may cost them more time to understand the program.
However, our findings are from the comparison of debugging time in three groups, while the absolute total debugging time will not affect.
On the other hand, to mitigate the bias from grouping, we randomly grouped participants according to their programming experience, and made each group of participants have close debugging capability in terms of developing years, i.e., one average about 6 years of developing experience each group. Besides, the participants from two groups finished four tasks for in debugging setting, which can alleviate the threats from participants.


The \textit{external} threat to validity lies in 
the four tasks used in user study, which may cause the findings may be not generalizable to all other cases. To mitigate this threat, we selected the tasks from different projects, covering different types. Additionally, we evenly distribute participants over different projects, which can mitigate the impact of projects to debugging process.

\section{Related Work} \label{section:relatedwork}
\subsection{Interactive Repair}
Some recent studies tried to involve developers in the program repair process. Cashin et al.~\cite{DBLP:conf/kbse/CashinMWF19} proposed PATCHPART, which clusters a set of generated patches by program invariants. Patches in a cluster are likely to be all correct or all incorrect, such that the developers ideally need only examine one patch per cluster. Compared with their approach, our approach actively asks questions rather than clusters the patches. We also present an empirical study showing that this process could increase the repair performance of developers.


B{\"{o}}hme et al.~\cite{DBLP:journals/corr/abs-1912-07758} introduced LEARN2FIX, which queries the developers to build a test oracle before patch generation to overcome overfitting. Compared with them, our study focuses on the patch review process after patch generation, and give evidences that our approach boosts the overall repair performance of developers.

\subsection{Interactive Debugging}
A lot of interactive debugging techniques~\cite{algorithmic-debugging-survey,DBLP:conf/icse/LiZdO18,DBLP:conf/icse/LinSXLD17,DBLP:conf/icse/KoM08,DBLP:conf/chi/KoM09,DBLP:conf/chi/KoM04,DBLP:conf/icsm/GongLJZ12} leverage user feedback to localize fault.
Algorithm debugging~\cite{algorithmic-debugging-survey} initially was proposed to resolve the functional programming debugging problem. Algorithm debugging first builds a debugging tree to reflect the method invocation for a failed test. Then, it repeatedly asks the developer to answer whether the input-output is correct for a method invocation in the debugging tree, and prunes the tree based on the answer until the fault is localized. 
Li et al.~\cite{DBLP:conf/icse/LiZdO18} improved algorithm debugging by leveraging spectrum based fault localization (SBFL) and dynamic dependences to decide the order of method invocations to be questioned. Similarly, Gong et al.~\cite{DBLP:conf/icsm/GongLJZ12} proposed an interactive fault localization technique to improve SBFL by asking developers to label the statements as faulty or clean, and updating the suspicious statement list.

Ko and Myers proposed Whyline~\cite{DBLP:conf/icse/KoM08,DBLP:conf/chi/KoM09,DBLP:conf/chi/KoM04}, which first records the program execution trace and allows developers to select some questions about program output. 
Whyline can give possible explanations according to the dynamic slicing until the developer finds the root cause of the fault.
Lin et al. proposed Microbat~\cite{DBLP:conf/icse/LinSXLD17} to improve Whyline by allowing developers to select the trace execution pattern, such as \textit{Correct Step, Wrong Variable Value}, and using developer's feedback to recommend some suspicious trace.

Different from these techniques, our approach aims to help patch review by interactively patch filtering, instead of improving the fault localization. Besides, our approach provided the differences among patched programs, while others do not.

\subsection{Patch Correctness Identification}
Since the weak test suites, a lot of researches focus on automatically identifying the correctness of patches.
Some approaches adopt a deterministic way. 
Xin and Reiss~\cite{Xin17} assume that there is an oracle, which can give the corresponding output for a newly generated input. If the output produced by the patched program violates the output produced by the oracle, the patch is incorrect. Yang et al.~\cite{Yang17} identify the correctness of patches by generating new test cases, which can obviously validate the oracle, such as crash and memory leak problems. Similarly, Gao et al.~\cite{DBLP:conf/issta/GaoMR19} use crash-freedom as the oracle to discard patches, which crash on the new tests.

Other approaches adopt a heuristic way to identify correctness of patches. 
Tan et al.~\cite{tananti} use anti-patterns to discard the patches which accord with the pre-defined patterns. 
Xiong et al.~\cite{DBLP:conf/icse/XiongLZ0018} determine the patches correctness by the behavior similarity of test case executions. 
Besides, Yu et al.~\cite{Yu17} indicate that test case generation can filter a part of incorrect patches, but cannot turn incorrect patches into correct ones.

Different from these approaches, our approach identifies the patch correctness by the developers. We adapted an interaction with the developers by asking questions related to the attributes of different patched programs. The developers can understand the bug and figure out the correct patches through the interaction. 

\subsection{Effect of Patches}
Tao et al.~\cite{DBLP:conf/sigsoft/TaoKKX14} have investigated the effect of automatic patch generation in realistic debugging scenarios. They observed that high-quality patches significantly improve debugging correctness while low-quality patches influence participants’ debugging correctness. Our user study also has a similar observation that a small number of high quality candidate patches could improve the performance of manual debugging.
 
Cambronero et al.~\cite{DBLP:conf/vl/Cambronero0CGR19} conducted a similar experiment except that the developers were provided with five candidate patches. The results show that when given candidate patches, the efficiency and correctness of developers did not be improved. 
Different from their observation, our study found that providing a large number of low quality patches would affect the performance of developers.

\section{Conclusion}\label{section:conclusion}
In this paper, we proposed an interactive patch filtering approach, which contains a two-stage algorithm, to provide tool support for patch review. We also implemented our approach as an Eclipse plugin called \toolname. The evaluation results show that our tool would significantly boost the repair performance of developers when the patch set contains high-quality patches, and would not significantly reduce the repair performance even when the patches are all of low-quality. These findings give many implications to future repair tool building, for example, (1) the repair tool of low precision can still useful, as the interactive process helps filter out incorrect patches without significantly affecting the repair performance; (2) a generated patch does not have to be fully correct to be useful, as incorrect patches could also provide good hints such as repair location and partial repair to the developer.


\bibliographystyle{ACM-Reference-Format}
\bibliography{ref,plde}

\end{document}